\documentstyle[epsfig,preprint,aps]{revtex}
\begin{document}
\draft
\def\bra#1{{\langle #1{\left| \right.}}}
\def\ket#1{{{\left.\right|} #1\rangle}}
\def\bfgreek#1{ \mbox{\boldmath$#1$}}
\title{$Q^2$--Dependence of the Gerasimov-Drell-Hearn Sum Rule}
\author{W. X. Ma, D. H. Lu, and A. W. Thomas}
\address{Department of Physics and Mathematical Physics\break
 and
        Special Research Centre for the Subatomic Structure of Matter,\break
        University of Adelaide, Australia 5005}
\author{Z. P. Li}
\address{Department of Physics, Peking University, \break 
Beijing 100871, P. R. China}
\maketitle
\vspace{-9.3cm}
\hfill ADP-97-46/T274
\vspace{9.0cm}

\begin{abstract}
We test the Gerasimov-Drell-Hearn (GDH) sum rule numerically by calculating 
the total photon absorption cross sections $\sigma_{1/2}$ and $\sigma_{3/2}$ 
on the nucleon via photon excitation of baryon resonances in the 
constituent quark model. 
A total of seventeen, low-lying, non-strange baryon resonances are included 
in this calculation. 
The transverse and longitudinal interference cross section, 
$\sigma_{1/2}^{TL}$,
is found to play an important role in the study of 
the  $Q^2$ variation of the sum rule. 
The results show that the GDH sum rule is saturated by these resonances 
at a confidence level of 94\%.
In particular, the $P_{33}(1232)$ excitation largely saturates the  
sum rule at $Q^2 = 0$, and dominates at small $Q^2$. 
The GDH integral has a strong 
$Q^2$-dependence below $Q^2= 1.0 \mbox{ GeV}^2$ and changes its sign 
around $Q^2= 0.3 \mbox{ GeV}^2$. 
It becomes weakly $Q^2$-dependent for
$Q^2 > 1.0 \mbox{ GeV}^2$ because of the quick decline
of the resonance contributions. 
We point out that the $Q^2$ variation of the 
GDH sum rule is very important for understanding the nucleon spin structure 
in the non-perturbative QCD region.
\end{abstract}

\section{Introduction}
The GDH sum rule connects the helicity structure of the total 
photo-absorption cross section 
to the ground state properties of the target  nucleon. 
Based on general principles such as  Lorentz invariance, 
electromagnetic gauge invariance, crossing symmetry,
causality, unitarity, and the less sound assumption that one can use an 
unsubtracted dispersion relation 
for the spin-dependent part of the forward Compton scattering amplitude,
the GDH sum rule can be written as\cite{GDH}
\begin{equation}
\int_{\nu_{thr}}^\infty {d\nu\over\nu} [\sigma_{1/2}(\nu,Q^2=0)
-\sigma_{3/2}(\nu,Q^2=0)] 
= -{2\pi^2\alpha\over M^2} \kappa^2.
\end{equation}
Here $\nu$ is the photon energy in the laboratory frame, 
$Q^2$ (= $-q^2$) is the
momentum transfer squared (zero for a real photon),
$\sigma_{1/2}$ and $\sigma_{3/2}$  are the
total photoabsorption cross sections for the total helicity 1/2 and 3/2 cases,
$M$ is the nuclon mass, $\alpha$ the hyperfine structure constant
and $\kappa$  the anomalous magnetic moment of the target nucleon. 
On the other hand, the analysis of the EMC experimental data (polarized
muon deep inelastic scattering)
suggests\cite{EMC}
\begin{equation}
\int_{\nu_{thr}}^\infty {d\nu\over\nu} [\sigma_{1/2}(\nu,Q^2)
-\sigma_{3/2}(\nu,Q^2)] 
= {0.2\over Q^2} \left(8\pi^2\alpha\over M^2\right) 
\end{equation}
in the region around $Q^2=10\mbox{ GeV}^2$. 
A comparison of Eq.~(1) and Eq.~(2) 
indicates that satisfying the GDH sum rule requires a dramatic change 
in the helicity structure of the photon-nucleon coupling between 
the real photon limit ($Q^2=0$) and the deep inelastic region.
  
This apparent sign change of the sum rule integral signifies a change 
of reaction mechanism between the deep inelastic scattering regime and 
the region of baryon resonance excitation at low and medium energies. 
This sharp $Q^2$ dependence of the GDH sum rule
provides an 
important consistency check for quark based hadronic models
and our deep understanding of the structure of hadrons. 
The idea that low lying resonances might saturate the GDH sum rule
was discussed many years ago\cite{Close72} and has
been investigated more quantitatively by Burkert and Li\cite{Burkert93}.  
On the experimental side, however, the sum rule has never been directly
measured  because  the technical
developments of the necessary polarized beam and target were not available.
However there are now a number of new experiments  
planned or underway at facilities like ELSA\cite{ELSA},
GRAAL\cite{GRAAL}, LEGS\cite{LEGS}, and MAMI\cite{MAMI}.

Since the sum rule is derived at zero $Q^2$, the baryon resonance 
excitation mechanism dominates the photon-nucleon coupling. 
Of particular interest is the question of whether the sum rule is 
truly saturated by low-lying baryon resonances. 
To  what extent do they saturate and  which resonance contributes the most?
Answers to these questions will be very interesting with regard to 
the interplay of nuclear and particle physics.
In this article, we study the GDH sum rule in the constituent quark model. 
We briefly present the GDH sum rule and our main physics 
ideas in section 2, followed by  
the necessary formalism for the helicity amplitudes in section 3. 
In section 4, we show our results and give a few concluding remarks.
  
\section{The Generalized  GDH sum rule}

To understand  the GDH sum rule, we start with a brief derivation 
considering the forward Compton scattering amplitude of a real 
photon on a nucleon. 
The corresponding scattering amplitude can be written out between the 
initial and 
final nucleon Pauli spinor $\chi_i$ and $\chi_f$,
\begin{equation}
T(\nu, \theta=0) = \chi_f^\dagger 
 [\vec{\epsilon^*_f} \cdot\vec{\epsilon}_i f(\nu) + 
 i\vec{\sigma} \cdot(\vec{\epsilon^*_f} \times\vec{\epsilon}_i) g(\nu)] \chi_i,
\end{equation}
where $f(\nu)$ and $g(\nu)$ are spin non-flip and flip amplitudes, 
$\vec{\epsilon}_f$ and $\vec{\epsilon}_i$ denote the polarization 
vectors of the 
initial and final photon and $\vec{\sigma}$ is the spin of the nucleon.
Both amplitudes, $f(\nu)$ and $g(\nu)$, are functions of the 
photon energy $\nu$ and 
can be expanded into a power series for small $\nu$,
where the  leading terms are determined by low energy theorems 
which are based only on Lorentz and gauge invariance\cite{Low54}.
\begin{eqnarray}
f(\nu)&=& -{e^2\over m} + (\alpha_N + \beta_N)\nu^2 + O(\nu^4), \\
g(\nu)&=& -{e^2\kappa^2\over 2m^2}\nu + \gamma_N\nu^3 + O(\nu^5). \label{g}
\end{eqnarray}
Here the first term in $f(\nu)$  is the Thomson limit,
 and the next term  is the contribution of 
the scalar polarizabilities of the nucleon ($\alpha_N$ for electric and 
$\beta_N$ for magnetic). Similarly, the leading term of $g(\nu)$ 
is from the anomalous magnetic moment $\kappa$, 
the next order is the vector polarizability of the nucleon. 

The two independent amplitudes, $f(\nu)$ and $g(\nu)$, can be 
determined by an experiment using circularly polarized photons and 
a polarized nucleon target with spin parallel 
($J_z=3/2$) and antiparallel ($J_z=1/2$) to the photon momentum. 
The corresponding amplitudes may be expressed as
\begin{equation}
T_{3/2} = f - g, \hspace{3cm} T_{1/2} = f + g.
\end{equation}
The optical theorem (unitarity) relates the 
imaginary parts of these amplitudes 
to the corresponding total photoabsorption cross sections,
\begin{equation}
\mbox{Im}[T_{1/2,3/2}(\nu)] = {\nu\over 4\pi} \sigma_{1/2,3/2}(\nu).
\end{equation}
From crossing symmetry, $f(\nu)$ must be even and 
$g(\nu)$  odd
under the transformation, $\nu\rightarrow -\nu$.
Thus, on the basis of analyticity, unitarity and crossing symmetry, 
we can now write a dispersion relation for $g(\nu)$  
\begin{equation}
\mbox{Re}[g(\nu)] = {2\nu\over \pi}\mbox{ P}\int_{\nu_{thr}}^\infty 
{d\nu'\over \nu'^2-\nu^2}
{\nu'\over 4\pi} {\sigma_{1/2}-\sigma_{3/2}\over 2},
\label{dispersion}
\end{equation}
The use of an unsubtracted dispersion relation relies on the additional,
reasonable hypothesis that 
$|g(\nu)|\rightarrow 0$ as $\nu\rightarrow\infty$.
Since the threshold energy is of the order of the pion mass,
this expression may be expanded as a power series in $\nu$ as well. 
Comparing the resulting series with the low energy expansion, Eq.~(\ref{g}), 
we can easily obtain the GDH sum rule given in Eq.(1).
Similarly, by taking the third derivative 
of Re[$g(\nu)$] we obtain the Burkardt-Cottingham (BC) sum rule, 
which relates
the helicity amplitudes to the nucleon vector polarizability\cite{BC70},
\begin{equation}
\int_{\nu_{thr}}^\infty {d\nu\over\nu^3} [\sigma_{1/2}(\nu,0)
-\sigma_{3/2}(\nu,0)] 
= {4\pi^2}\gamma_N.
\end{equation}

In QCD, the GDH sum rule provides an important constraint on the 
spin structure of the composite
nucleon. It is complementary to the important Bjorken\cite{Bjorken66} and 
Ellis-Jaffe\cite{EJ74} sum rules
which relate the first moment of the nucleon's first spin structure function, 
$g_1(x,Q^2)$, to the axial charge of the nucleon\cite{Bass97}.
It should be emphasized that the GDH sum rule in Eq.(1)  
is derived for a real photon which is transverse.
For electron-nucleon scattering (shown in Fig.~1), however, 
the exchanged photon must be  virtual and 
hence it can also be longitudinal  ($\lambda=0$). 
In this case, the additional interference term between transverse 
and longitudinal photons, $\sigma_{1/2}^{TL}$, 
also contributes to the $Q^2$ dependence 
of the GDH sum rule. 
Thus, one should not just consider the $Q^2$ dependence 
of the difference
($\sigma_{1/2}^T - \sigma_{3/2}^T$) alone 
in studying the $Q^2$ dependence of the GDH sum rule. 

As is well known, the double differential cross section for electron deep
inelastic scattering from a nucleon in one photon exchange approximation
(see Fig.~1 for notation and kinematics) can be written as the scalar product 
of a leptonic tensor
$L_{\mu\nu}$ and a hadronic tensor $W_{\mu\nu}$\cite{Leader}
\begin{equation}
{d^2\sigma\over d\Omega dE'} = {\alpha^2\over Q^4}{E'\over E}
L_{\mu\nu} W^{\mu\nu},
\end{equation}
with $Q^2=-(k_\mu-k'_\mu)^2$.
Both tensors can be decomposed
into a symmetric (S) and an antisymmetic part (A), so that
\begin{eqnarray}
L_{\mu\nu}&=&L_{\mu\nu}^{(S)} + i L_{\mu\nu}^{(A)}, \\ 
W_{\mu\nu}&=&W_{\mu\nu}^{(S)} + i W_{\mu\nu}^{(A)},
\end{eqnarray}
where the $L_{\mu\nu}^{(S)}$ and $ L_{\mu\nu}^{(A)}$ are completely known and
are given by
\begin{eqnarray}
L_{\mu\nu}^{(S)} &=& 2 [k_\mu k'_\nu + k'_\mu k_\nu - 
  g_{\mu\nu}(k\cdot k' - m_e^2)], \\
L_{\mu\nu}^{(A)} &=& 2\epsilon^{\mu\nu\lambda\sigma}q_\lambda \sigma_\sigma,
\end{eqnarray}
with $\sigma$  being the lepton polarization 
vector and $g_{\mu\nu}$ the metric tensor. 
The  unknown $W_{\mu\nu}$, which describes the internal 
structure of the nucleon, depends on non-perturbative QCD dynamics. 
In principle, it can be expressed as follows,
\begin{eqnarray}
W_{\mu\nu}^{(S)} &=& ({q_\mu q_\nu\over q^2}-g_{\mu\nu}) W_1(\nu,Q^2)
 + {1\over M^2}(p_\mu - {p\cdot q\over q^2} q_\mu)
               (p_\nu - {p\cdot q\over q^2} q_\nu) W_2(\nu,Q^2), \\
W_{\mu\nu}^{(A)} &=& \epsilon_{\mu\nu\lambda\sigma} q^\lambda 
\{ s^\sigma M [G_1(\nu,Q^2) + {p\cdot q \over M^2} G_2(\nu,Q^2)] 
 - p^\sigma {s\cdot q \over M} G_2(\nu,Q^2) \}, 
\end{eqnarray}
where $s$ is the nucleon polarization vector, 
the $W_{1,2}(\nu,Q^)$ and $G_{1,2}(\nu,Q^2)$ refer  
to unpolarized and polarized structure functions (scalar functions), 
respectively.
The functions $W_{1,2}(\nu,Q^2)$, which  have played a seminal role in
the development of our current understanding of hadron structure,
 carry information about the overall 
distribution (density) of quarks and gluons in the nucleon.
The function $G_{1}(\nu,Q^2)$ probes the spin distribution of quarks 
in a polarized nucleon, while
$G_2(\nu,Q^2)$ involves higher twist contributions 
and its physical interpretation is obscure.

The total virtual photoabsorption cross section can then be expressed 
in terms of the hadronic tensor $W_{\mu\nu}$\cite{Ioffe}
\begin{equation}
\sigma_{\lambda}^{\gamma^*N} = 
{4\pi^2\alpha\over K}\epsilon^{\mu*}(\lambda)
W_{\mu\nu}(\nu,Q^2) \epsilon^\nu(\lambda),
\end{equation} 
where $K=\sqrt{\nu^2 + Q^2}$ is the flux of virtual photons 
(using the Gilman convention\cite{Gilman68}) and 
$\epsilon^\nu(\lambda)$ is the polarization vector
of the virtual photon.
Among these different $\sigma_{\lambda}^{\gamma^*N}$, 
only four cross sections are 
independent under rotational, parity and time reversal invariance. 
They are labeled by the photon polarization $T$ (transverse) 
and $L$ (longitudinal),
and can be explicitly expressed in terms of the structure functions, 
\begin{eqnarray}
\sigma_{3/2}^T &=& {4\pi^2\alpha\over K} 
          [W_1(\nu,Q^2) - M\nu G_1(\nu,Q^2) + Q^2 G_2(\nu,Q^2)], \label{Da} \\
\sigma_{1/2}^T &=& {4\pi^2\alpha\over K} 
          [W_1(\nu,Q^2) + M\nu G_1(\nu,Q^2) - Q^2 G_2(\nu,Q^2)], \label{Db} \\
\sigma_{1/2}^L &=& {4\pi^2\alpha\over K} 
          [W_2(\nu,Q^2)(1 + {\nu^2\over Q^2}) - W_1(\nu,Q^2)], \label{Dc} \\
\sigma_{1/2}^{TL} &=& {4\pi^2\alpha\over K}\sqrt{Q^2} 
          [M G_1(\nu,Q^2) + \nu G_2(\nu,Q^2)]. \label{Dd}
\end{eqnarray}
These four independent cross sections
contain all the information about the hadron structure 
governed by the underlying strong interaction theory.

Combining Eqs.~(\ref{Da}), (\ref{Db}) and (\ref{Dd}) 
the spin dependent structure function $G_1(\nu,Q^2)$
can now be expressed in terms of three total absorption cross sections 
$\sigma_{1/2}^T$, $\sigma_{3/2}^T$, and $\sigma_{1/2}^{TL}$,
\begin{equation}
G_1(\nu,Q^2) = {1\over 8\pi^2 \alpha M} {\nu\over\sqrt{\nu^2+Q^2}}
[\sigma_{1/2}^T - \sigma_{3/2}^T + {2\sqrt{Q^2}\over\nu}\sigma^{TL}_{1/2}].
\label{G1}
\end{equation}
To investigate the $Q^2$--dependence of the GDH sum rule, 
it is convenient to define
the GDH integral\cite{Leader,Ioffe}
\begin{equation}
I(Q^2) = M \int_{Q^2/2M}^\infty {d\nu\over \nu} \, G_1(\nu,Q^2).
\end{equation}
It is clear that the $\sigma^{TL}_{1/2}$ term vanishes at $Q^2=0$
so the original GDH sum rule for a real photon can be recovered.
In the perturbative region, this term also vanishes.
Thus, $I(Q^2)$ 
is a generalization for the GDH sum rule suitable to study the 
 $Q^2$-dependence of the sum rule. 
 
\section{Helicity amplitudes in the constituent quark model}

As is  well known, in the small $Q^2$ region the dominant mechanism
for  virtual photon absorption is through photo-excitation
of baryon resonances. 
Therefore, the total photoabsorption cross sections, 
$\sigma_{1/2}^T$, $\sigma_{3/2}^T$ and $\sigma_{1/2}^{TL}$, 
can be calculated in the constituent quark model which has been 
quite successful in hadronic physics. In the following subsection,
we shall express the total absorption cross sections in terms of helicity
amplitudes $A_{1/2}$, $A_{3/2}$ and $S_{1/2}$.

\subsection{Helicity amplitudes and electromagnetic interaction}
The total cross sections for the electromagnetic excitation of a given 
resonance from the nucleon can be 
completely specified by the corresponding helicity amplitudes 
$A_{1/2}$, $A_{3/2}$ and $S_{1/2}$. As illustrated in Fig.~2. 
these helicity amplitudes are defined as follows
\begin{eqnarray}
A_{1/2} &=& \bra{^d[\underline{B}]_J,[\underline{A},L^P]_N,S_z=1/2} H_t^{em} 
 \ket{^2[\underline{8}]_{1/2},[\underline{56},0^+]_0,S_z=-1/2}, \label{A12}\\
A_{3/2} &=& \bra{^d[\underline{B}]_J,[\underline{A},L^P]_N,S_z=3/2} H_t^{em} 
 \ket{^2[\underline{8}]_{1/2},[\underline{56},0^+]_0,S_z=1/2}, \label{A32}
\end{eqnarray}
and
\begin{equation}
S_{1/2} = \bra{^d[\underline{B}]_J,[\underline{A},L^P]_N,S_z=1/2} H_l^{em} 
 \ket{^2[\underline{8}]_{1/2},[\underline{56},0^+]_0,S_z=1/2}, \label{S12}
\end{equation}
where $\ket{^d[\underline{B}]_J,[\underline{A},L^P]_N,S_z}$ 
stands for the spin-flavor-space
part of the baryon wave function 
in the standard $[SU_{\rm SF}(6)\otimes O(3)]_{\rm sym} \otimes SU_c(3)$ 
classification scheme.
The $[\underline{A},L^P]_N$ are the $SU_{\rm SF}(6)$ multiplets with 
$\underline{A}$ = $\underline{56}$, $\underline{70}$, or $\underline{20}$, 
$L$  is the total orbital angular momentum, $N$ the total quantum number
of excitations of all the modes and $P=(-1)^N$ is the parity of the state.
The $^d[\underline{B}]_J$ specifies the $SU_F(3)$ multiplet states, 
with \underline{B}=\underline{1}, \underline{8}, or \underline{10} 
(singlet, octet, and decuplet), $d$  the spin multiplicity and 
$J$ the total
angular momentum of the state. The proton in the initial state is 
thus denoted by 
$\ket{^2[\underline{8}]_{1/2},[\underline{56},0^+]_0,S_z}$,
 with $S_z=\pm 1/2$. 

The $ H_t^{em}$ and $H_l^{em}$  are the transverse and 
longitudinal transition operators, respectively, and 
\begin{equation}
H_t^{em}=H_{NR} +H_{SO} + H_{NA}
\end{equation}
with
\begin{eqnarray}
H_{NR} &=& -\sum_{i=1}^3 [{e_i\over 2 m_i} (\vec{p}_i\cdot\vec{A}_i + 
\vec{A}_i\cdot\vec{p}_i) + \mu_i\vec{\sigma}_i\cdot\vec{B}_i - e_i\phi_i ], \\
H_{SO} &=& -{1\over 2}\sum_{i=1}^3 \left[ 2\mu_i-{e_i\over 2 m_i}\right]
{\vec{\sigma}_i\over 2m_i}(\vec{E}_i\times\vec{p}_i-\vec{p}_i\times\vec{E}_i), \\
H_{NA} &=& \sum_{i<j} {1\over 4M_T}({\vec{\sigma}_i\over m_i}-
{\vec{\sigma}_j\over m_j})(e_j \vec{E}_j\times\vec{p}_i- 
e_i \vec{E}_i\times\vec{p}_j), \label{NA}
\end{eqnarray}
where the subscript $i$ refers to $i$-th quark, and the three components of
the Hamiltonian correspond to the nonrelativistic (NR),  spin-orbit (SO),
 and non-additive (NA) interactions, respectively. 
The longitudinal transition operator, $H_l^{em}$, is defined as 
\begin{equation}
H_l^{em} = \epsilon_0 J_0 - \epsilon_3 J_3,
\end{equation}
where $\epsilon_\mu = {1\over\sqrt{Q^2}}\{k_3,0,0,\nu\}$
is the photon longitudinal polarization vector and
$J_\mu$ the nucleon electromagnetic current. 
Using the gauge invariance condition 
$k_\mu J^\mu = k_\mu \epsilon^\mu = 0$, one obtains 
$\bra{\psi_f}H_l^{em} \ket{\psi_i}=
{\sqrt{Q^2}\over k_3}\bra{\psi_f}J_0\ket{\psi_i}$.
Therefore, the matrix element $\bra{\psi_f}H_l^{em}\ket{\psi_i}$ 
does not depend on
the choice of the current as long as it is gauge invariant. 
It is noted that the longitudinal transition is proportional to $\sqrt{Q^2}$
 and so  vanishes in the real photon limit. 
In terms of quarks, the zeroth component of the longitudinal
transition operator is expressed by 
\begin{eqnarray}
J_0^{em} &=& { 1\over \sqrt{2\nu} }
 \{\sum_j \left[e_j + {i e_j\over 4m_j^2}
\vec{k}\cdot (\vec{\sigma}_j\times\vec{p}_j\right] e^{i\vec{k}\cdot\vec{r}_j}
  \nonumber \\
 &-& 
 \sum_{j<i}{i\over 4M_T} 
\left( {\vec{\sigma}_j\over m_j}-{\vec{\sigma}_i\over m_i}\right)
[e_j(\vec{k}\times\vec{p}_i) e^{i\vec{k}\cdot\vec{r}_j}
-e_i(\vec{k}\times\vec{p}_j) e^{i\vec{k}\cdot\vec{r}_i}]\}. \label{J0}
\end{eqnarray}
The $M_T$ in Eq.~(\ref{NA}) and (\ref{J0}) is the total mass of the quark
system.
The first term in Eq.~(\ref{J0}) is the charge operator used in the 
conventional calculation of the longitudinal helicity amplitudes, 
while the second 
and third terms are the spin-orbit and non-additive terms which have 
counterparts in the transverse electromagnetic transition and represent the
relativistic corrections to the first term of order $O(v^2/c^2)$.
In Ref.~\cite{Li90} it was  pointed out that the spin-orbit and non-additive
terms are crucial in reproducing the experimental data.
However, most studies have neglected these two terms.
In this work, we include 
these two terms in our calculations.
Furthermore, we note  that the Hamiltonian used here satisfies the low energy
theorem\cite{Low54}. The details of our calculations 
for these helicity amplitudes,
$A_{1/2}$, $A_{3/2}$ and $S_{1/2}$, are given 
in the previous papers\cite{Li93}.

\subsection{The total absorption cross sections}
Given $A_{1/2}$, $A_{3/2}$ and $S_{1/2}$ for each resonance, 
the total virtual photon absorption cross sections, 
$\sigma_{1/2}^T$, $\sigma_{3/2}^T$, and $\sigma_{1/2}^{TL}$, 
can be easily obtained. 
Using a Breit-Wigner parameterization for each resonance,
the  total photoabsorption cross sections
are the incoherent sums of the contributions from the 
 individual baryon resonances\cite{FFR65}, 
\begin{eqnarray}
\sigma_{1/2,3/2}^T(\nu,Q^2) &=& \sum_R \left( {2M\over W+W_R}\right)
{\Gamma_R\over (W-W_R)^2 + \Gamma_R^2/4} 
\left| A_{1/2,3/2}^R(\nu,Q^2)\right|^2, \\
\sigma_{1/2}^{TL}(\nu,Q^2) &=& \sum_R \left( {2M\over W+W_R}\right)
{\Gamma_R\over (W-W_R)^2 + \Gamma_R^2/4} 
[A_{1/2}^{R*} S_{1/2}^R +  S_{1/2}^{R*} A_{1/2}^R]
\end{eqnarray}
where $W$ is the center-of-mass energy, $W^2=(p+q)^2$,
$W_R$ and $\Gamma_R$ are  the mass and the total width of the resonance $R$, 
and $A_{1/2,3/2}^R$ and $S_{1/2}^R$ correspond to the helicity amplitudes 
for the  
transverse and longitudinal photon to produce resonance $R$, 
respectively.

\section{Results and concluding remarks}
The calculation of the helicity amplitudes $A_{1/2}^R$, $A_{3/2}^R$ 
and $S_{1/2}^R$
has been carried out for 17 baryon resonances:
$P_{33}(1232)$, $P_{11}(1470)$, 
$D_{13}(1520)$, $S_{11}(1535)$, $S_{11}(1650)$, $S_{31}(1620)$, $D_{15}(1675)$,
$D_{33}(1675)$, $D_{13}(1720)$,
$F_{15}(1680)$, $P_{13}(1720)$, $P_{31}(1920)$, $P_{35}(1905)$, $P_{33}(1920)$,
$P_{37}(1950)$, 
$P_{11}(1705)$,  and $P_{33}(1600)$.
Non-relativistic harmonic oscillator wave functions,
which have  been shown  to reproduce the well-known 
properties of these baryon resonances quite successfully, 
are used in the 
calculation. The helicity amplitudes in the constituent quark model
are consistent with the experimental data for the first few low-lying 
resonances. For further details of the calculation, 
we refer to Ref.~\cite{Li93}.

To compare with the experimental cross sections, the non-resonant background 
contributions to the same processes have to be considered.
Here we use a phenomenological method which ignores
possible interference between background and resonances in the same 
multipole channel\cite{Haiden79}. 
Our predicted $Q^2$ dependence of the GDH sum rule is shown in Fig.~3.
The solid curve represents the full contributions 
from all 17 resonances included, whereas
the dash-dotted curve gives the $P_{33}(1232)$ contribution alone.
Clearly, the bulk features of the $Q^2$ dependence come from
the dominant $P_{33}(1232)$ resonance for small momentum transfers. 
However, the contributions from
higher resonances are sizable especially after the sign-flip point 
($Q^2\sim 0.3\mbox{ GeV}^2$) where the $P_{33}(1232)$ contribution nearly
vanishes.
In the same figure, 
we also show  the  pQCD $Q^2$--evolution [see Eq.~(2)]  
with the dashed line. There is a big gap in the small $Q^2$ region. 

(i) The GDH sum rule is saturated by the contributions from baryon
resonances at a confidence level of 94\%. The predicted value
of the GDH integral, $I(Q^2=0)$, is $0.492 \mbox{ GeV}^{-2}$ 
which is to be compared with 
${2\pi^2\alpha\over M^2}\kappa^2 = 0.524 \mbox{ GeV}^{-2}$ for the proton.

(ii) The GDH sum rule has a strong $Q^2$-dependence below 
$Q^2=1.0 \mbox{ GeV}^2$, and a sign flip at about $Q^2= 0.3 \mbox{ GeV}^2$.
This is in agreement with the recent data from the E143 collaboration
at SLAC which indicates that the sign flip point should be even smaller than
$0.5 \mbox{ GeV}^2$. 
A weak $Q^2$ dependence is predicted in the region of $Q^2$ 
larger than $1.0 \mbox{ GeV}^2$. This $Q^2$-dependence comes mainly
from the contribution of the $P_{33}(1232)$ resonance.
 At small momentum transfers, the main portion (roughly 75\%) of the GDH sum 
rule integral is provided by the contribution of $P_{33}(1232)$ 
photoproduction.
This contribution is negative at small $Q^2$, turns positive
around $Q^2=0.3 \mbox{ GeV}^2$ and soon  decreases rapidly with a further
increase of $Q^2$. Our interpretation is the following: 
at small $Q^2$, the magnetic dipole
transitions dominate $P_{33}(1232)$ photoproduction and 
$\sigma_{1/2}/\sigma_{3/2}\sim 3$. At large $Q^2$, because of chiral 
invariance, only $\sigma_{1/2}$
survives and this leads to a sign change in the GDH integral.
For larger $Q^2$, the form factor for the proton to $P_{33}(1232)$ transition 
starts to act and quickly makes the $P_{33}(1232)$ contribution 
negligible after $Q^2\sim 1 \mbox{ GeV}^2$.

(iii) Our analysis also shows that the behavior in the deep inelastic
region cannot be naively extended to small $Q^2$ region where
the resonances dominate
(see the dash-dotted line in Fig.~3). Qualitative agreement with the 
extrapolated EMC analysis at moderate $Q^2$ may be achieved 
only when the dominant contribution
from $P_{33}(1232)$ resonance is neglected. One cannot use pQCD 
in the small $Q^2$ region.

(iv) The calculation also shows that the contribution from $\sigma_{1/2}^{TL}$
cannot be neglected in the small $Q^2$ region. The longitudinal transitions
of baryon resonances play an important role in the understanding of  
baryon structure. In particular, the longitudinal transition 
amplitude $S_{1/2}(Q^2)$ decreases as $Q^2$ increases and  
it is therefore important
to consider the $S_{1/2}(Q^2)$ contribution in the small $Q^2$ region, 
especially for the transitions
between the nucleon and the $P$-wave resonances, which are accessible in the 
experiments at TJNAF.

In conclusion, we have tested the GDH sum rule numerically 
by studying inelastic
electron-proton scattering in the constituent quark model.
The virtual photon coupling to the nucleon was assumed to occur
mainly through the 
excitation of baryon resonances where the helicity amplitudes
could be calculated in the constituent quark model. 
The electromagnetic interaction Hamiltonian which we used
includes spin-orbit and non-additive terms in addition to the usual
nonrelativistic pieces.  
Thus the all important relativistic effects at medium and
high energy were effectively included.
Note that the interference cross section,
$\sigma_{1/2}^{TL}$, which has often been omitted before,
is now included in our calculation. 

Of course, there are many aspects of this calculation which 
could, in principle,
be improved. For example, because of the tensor structure of the hyperfine 
interaction the total spin and the total orbital angular momentum are no longer
separately conserved, and this will induce  $SU(6)_{\rm SF}\otimes O(3)$ 
configuration mixing. The corresponding baryon states are therefore 
superpositions of the $SU(6)_{\rm SF}\otimes O(3)$ basis states, 
\begin{equation}
\ket{\Phi_{\rm baryon}} = \sum_i c_i 
\ket{\Phi^i_{SU(6)_{\rm SF}\otimes O(3)}}.
\end{equation}
This configuration mixing may change the prospective theoretical result.
It is worthwhile to recall that there are two assumptions in deriving the 
GDH sum rule. The first is the low energy theorem for forward Compton 
scattering, which is well accepted, and the second is the 
validity of the unsubtracted dispersion relation [see Eq.~(\ref{dispersion})]. 
The GDH sum rule itself is general and model independent. 
It is reasonably supported by the pion photoproduction data\cite{Karlinger73}
and should provide
an important constraint and test for models which describe electromagnetic
transitions of the nucleon. 

This work was partially supported by the Australian Research Council. 
W. X. Ma would like to acknowledge the kind hospitality of the CSSM.

\begin{figure}
\vspace{2.5cm}
\centering{\
\epsfig{file=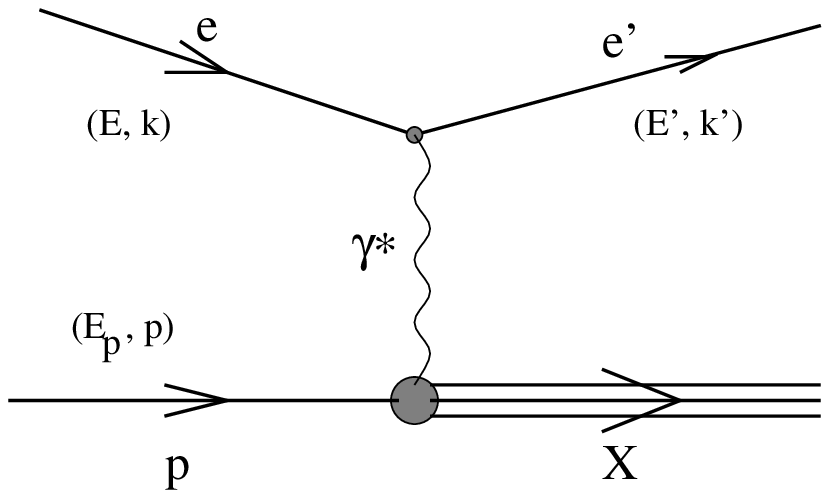,height=4.5cm,width=12cm}
\vspace*{1cm}
\vspace{1cm}
\caption{Schematic representation of electron deep inelastic scattering from
the nucleon. $\gamma^*$ stands for a virtual photon.}
\label{fig1.ps}}
\end{figure}

\newpage
\vspace*{1cm}
\begin{figure}
\centering{\
\epsfig{file=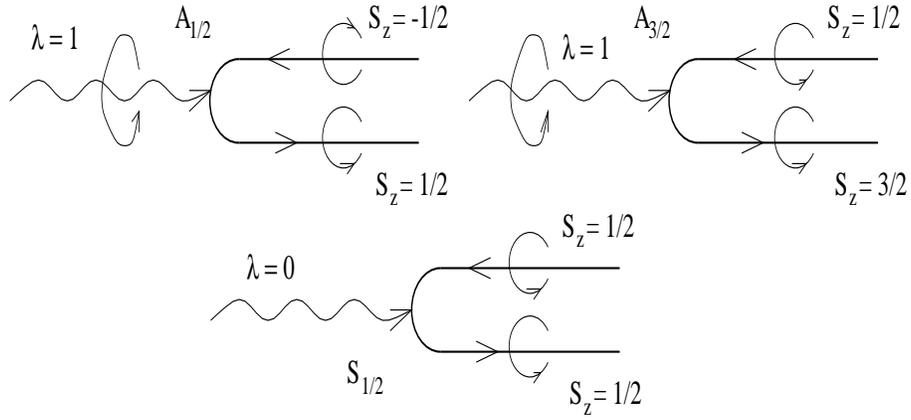,height=5.5cm,width=12cm}
\vspace{1cm}
\caption{Illustration of helicity amplitudes for a virtual photon scattering 
off a nucleon. The amplitudes $A_{1/2}$ and $A_{3/2}$ are 
for a transverse photon being absorbed by a nucleon with antiparallel 
and parallel  spin projections with respect to the photon, 
and $S_{1/2}$ corresponds to a longitudinal photon.}
\label{fig2.ps}}
\end{figure}

\newpage
\vspace*{1cm}
\begin{figure}
\centering{\
\epsfig{file=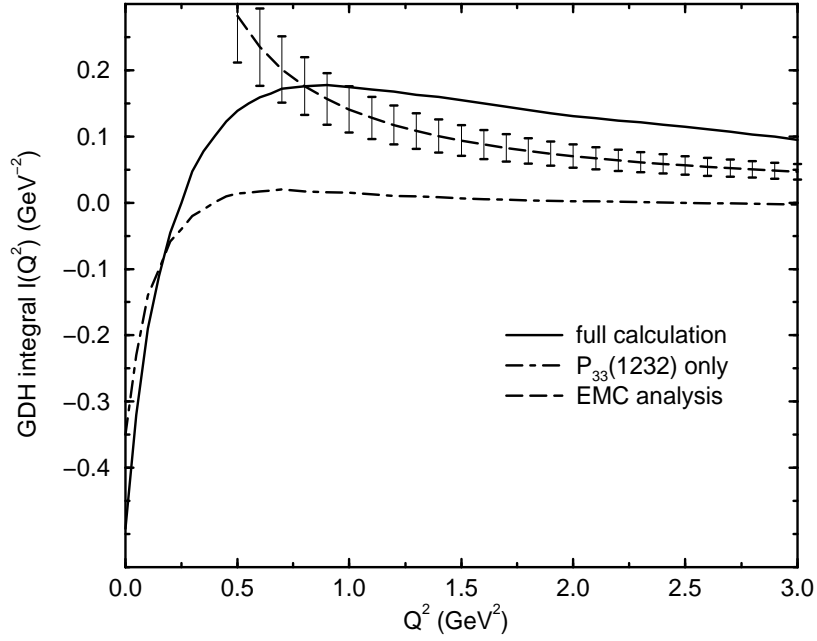,height=10cm,width=12cm}
\caption{$Q^2$-dependence of the GDH sum rule.
The solid curve represents total contributions from
17 baryon resonances while the dot-dashed curve gives
the $P_{33}(1232)$ contribution alone. The  dashed line with error bars
is from the pQCD analysis of the EMC experiment which is only valid 
for large momentum transfer.} 
\label{fig3.ps}}
\end{figure}

\end{document}